\documentclass[sigconf]{acmart}
\AtBeginDocument{%
  }

\usepackage{multirow}
\usepackage{multicol}
\usepackage{adjustbox}
\usepackage{makecell}
\usepackage{diagbox} 
\usepackage{enumitem}

\setcopyright{acmlicensed}
\copyrightyear{2025}
\acmYear{2025}
\acmDOI{XXXXXXX.XXXXXXX}
\acmConference[ICMR'25]{ACM Conference}{June 30 - July 3, 2025}{Chicago, USA}

\acmISBN{978-1-4503-XXXX-X/18/06}

\copyrightyear{2025}
\acmYear{2025}
\setcopyright{acmlicensed}\acmConference[ICMR '25]{Proceedings of the 2025 International Conference on Multimedia Retrieval}{June 30-July 3, 2025}{Chicago, IL, USA}
\acmBooktitle{Proceedings of the 2025 International Conference on Multimedia Retrieval (ICMR '25), June 30-July 3, 2025, Chicago, IL, USA}
\acmDOI{10.1145/3731715.3733483}
\acmISBN{979-8-4007-1877-9/2025/06}

\begin{document}

\title{Let Network Decide What to Learn: Symbolic Music Understanding Model Based on Large-scale Adversarial Pre-training}


\author{Zijian Zhao}
\affiliation{%
  \institution{School of Computer Science and Engineering \\ Sun Yat-sen University}
  \city{Guangzhou}
  \country{China}}
   \email{zhaozj28@mail2.sysu.edu.cn}
    \orcid{0000-0002-3326-9650}

\begin{abstract}
As a crucial aspect of Music Information Retrieval (MIR), Symbolic Music Understanding (SMU) has garnered significant attention for its potential to assist both musicians and enthusiasts in learning and creating music. Recently, pre-trained language models have been widely adopted in SMU due to the substantial similarities between symbolic music and natural language, as well as the ability of these models to leverage limited music data effectively. However, some studies have shown the common pre-trained methods like Mask Language Model (MLM) may introduce bias issues like racism discrimination in Natural Language Process (NLP) and affects the performance of downstream tasks, which also happens in SMU. This bias often arises when masked tokens cannot be inferred from their context, forcing the model to overfit the training set instead of generalizing.  To address this challenge, we propose Adversarial-MidiBERT for SMU, which adaptively determines what to mask during MLM via a masker network, rather than employing random masking. By avoiding the masking of tokens that are difficult to infer from context, our model is better equipped to capture contextual structures and relationships, rather than merely conforming to the training data distribution. We evaluate our method across four SMU tasks, and our approach demonstrates excellent performance in all cases. The code for our model is publicly available at \href{https://github.com/RS2002/Adversarial-MidiBERT}{https://github.com/RS2002/Adversarial-MidiBERT}.


\end{abstract}





\maketitle

\section{Introduction}

Music Information Retrieval (MIR) plays a crucial role in various applications, including recommendation systems in music apps and AI agents for music creation. With advancements in computer music, symbolic music—representing music through a structured sequence of notes—has gained significant attention, as most contemporary music is initially created and recorded using symbolic formats like MIDI \cite{MIDI}. Symbolic Music Understanding (SMU) has emerged as a key research direction within MIR, aiming to assist both musicians and enthusiasts in learning, teaching, and creating music.

Given the similarities between symbolic music and natural language, language models have been extensively utilized in SMU. For instance, the Bidirectional Encoder Representations from Transformers (BERT) \cite{BERT} model has demonstrated promising performance in SMU \cite{MusicBERT,MidiBERT}. A critical factor contributing to the success of current language models, particularly Large Language Models (LLMs), is their ability to leverage vast amounts of unlabeled data during pre-training to learn fundamental data structures and relationships. This pre-training mechanism has proven effective in domains with limited data, such as music \cite{PianoBART} and signals \cite{csibert}, thereby enhancing model performance in downstream tasks \cite{pretrain1}.

Among the prevalent pre-training methods for language models is the Mask Language Model (MLM) \cite{BERT}, particularly in encoder-only architectures. This method randomly masks certain tokens and trains the model to recover them. However, since some tokens cannot be inferred from the given context, training the model on such tasks can introduce bias issues, such as discrimination based on gender, age, or race in Natural Language Processing (NLP) \cite{bias2}. Research has indicated that these biases can significantly impair model performance in downstream tasks, including classification and generation \cite{bias1}. Consequently, even in the absence of ethical discrimination issues in music, bias can adversely affect model performance in SMU.

Before delving deeper into the bias problem, we present a straightforward example from NLP that may be more accessible to most readers than a music-related example. Consider the original sentence, "She is good at math." If we mask it to read, "[MASK] is good at math.", the model can only recover the masked token based on the training data distribution for optimal accuracy, rather than considering the contextual relationship, as there is no information indicating the subject. When the training data distribution regarding gender in various contexts is imbalanced, the model may default to recovering the [MASK] as "He", resulting in gender bias. In this case, we hope the model learns the basic grammatical structure of phrases like "be good at", rather than inferring "she" and "math" solely based on data distribution to achieve higher accuracy. Similarly, in music, we aim for the model to learn common musical structures, relationships, and regularities—such as basic modes, riff patterns, and modulation techniques—rather than developing habits specific to particular composers. In this paper, we define tokens that can be inferred from context as context-dependent tokens and those that cannot as context-free tokens. We argue that only masking context-dependent tokens in MLM is beneficial, as training the model to recover context-free tokens may lead to overfitting to the training set.

However, most current methods addressing bias are limited to the field of NLP and are challenging to transfer to other domains. Many of these methods tackle specific bias issues, such as sexism \cite{bias_gender} and regional bias \cite{bias_region}, but do not apply similarly in other areas. For instance, data augmentation \cite{bias_aug} is a promising approach in NLP. However, due to the less pronounced bias issues in domains like music and signal processing, it remains unclear how to effectively clean, generate, or modify data without sufficient domain-specific knowledge. To address the bias problem in SMU, we propose Adversarial-MidiBERT: 





\begin{itemize}[noitemsep,leftmargin=*]
\item[$\bullet$]  We introduce an adversarial pre-training approach that uses a masker network to selectively mask context-dependent tokens during MLM, avoiding context-free tokens. This reduces bias from dataset-specific patterns, enabling the model to learn robust musical structures like modes and riffs, improving generalization across SMU tasks.
\item[$\bullet$]  We propose a mask fine-tuning method that applies random [MASK] tokens during fine-tuning to align pre-training and fine-tuning phases. This enhances model robustness and performance on downstream tasks by maintaining contextual inference capabilities, especially with limited music data.
\item[$\bullet$] Experimental results demonstrate that our method achieves excellent performance across four music understanding tasks, including composer classification, emotion classification, velocity prediction, and melody extraction. Ablation studies confirm the effectiveness of our adversarial and fine-tuning mechanisms, highlighting the model’s versatility for music analysis.
\end{itemize}


\subsection{Overview}
Similar to most BERT-based methods, our Adversarial-MidiBERT involves two phases: MLM pre-training and fine-tuning. During pre-training, we introduce an adversarial mechanism to prevent context-free tokens from being masked. Specifically, we train a masker network to select tokens for masking and a recoverer network to recover them. The masker aims to minimize the recovery accuracy of the recoverer, thereby favoring the selection of context-free tokens. These tokens can only be inferred from the dataset distribution, which does not guarantee precise recovery. It is likely that the recovery accuracy for context-free tokens will be lower than that for context-dependent tokens, as a well-trained recoverer can easily infer context-dependent tokens from the surrounding information. After several epochs, we can freeze the tokens with a high selection probability according to the masker, preventing them from being masked in subsequent training. This allows the recoverer to focus solely on learning from context-dependent tokens. During fine-tuning, we implement a mask fine-tuning mechanism, where random [MASK] tokens replace input tokens to reduce the gap between pre-training and fine-tuning, thereby improving model performance efficiently.

\begin{figure}
\centering 
\includegraphics[width=0.5\textwidth]{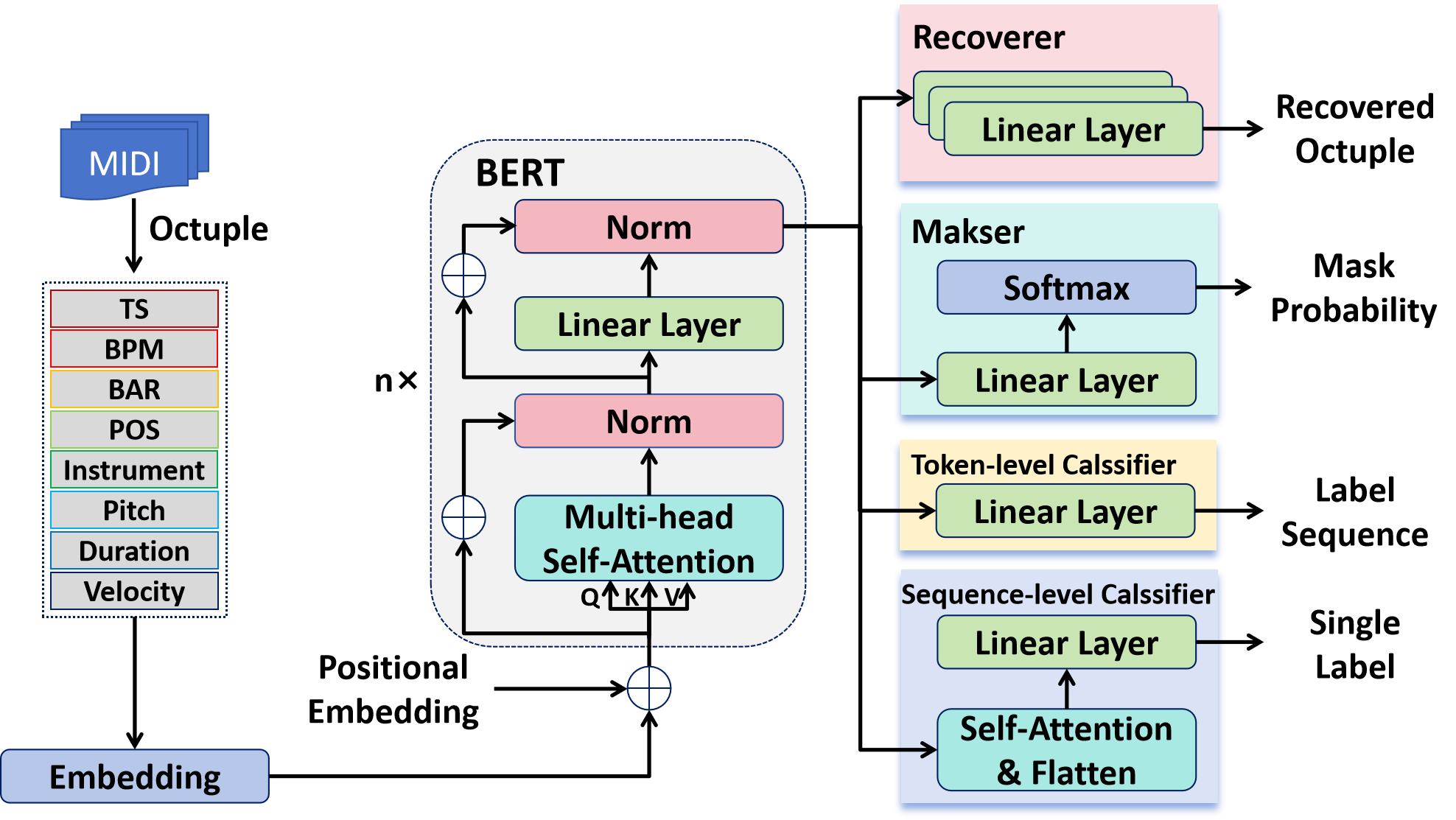}
\caption{Network Architecture}
\label{fig:main}
\end{figure}

Before delving deeper into the training mechanism, we first introduce our network and data structure. To embed MIDI music information, we employ Octuple \cite{MusicBERT} to represent the symbolic music structure. It transforms each MIDI file into a sequence of tokens, each representing one music note and possessing eight attributes: time signature (TS), tempo (BPM), bar position (BAR), relative position within each bar (POS), instrument, pitch, duration, and velocity. To input these tokens into the network, we utilize eight embedding layers to embed them individually, and then concatenate the embedding results. This process yields an embedding sequence suitable for input to the network, as shown in Fig. \ref{fig:main}. In this configuration, a BERT \cite{BERT} (the encoder component of the Transformer \cite{Transformer}) serves as the backbone, complemented by an embedding layer for music information and multiple output heads for different tasks. Specifically, the recoverer head addresses the MLM task by generating classification probabilities for each attribute at each position. The masker head generates the mask probability for each token. Additionally, the sequence-level and token-level classifiers are employed in downstream classification tasks, producing classification probabilities for the entire music sequence and each token, respectively. Due to page limitations, please refer to \cite{PianoBART,MusicBERT} for more details regarding the model structure and Octuple information.

\section{Proposed Method} \label{Proposed Method}

\begin{figure}[!t]
\centering 
\includegraphics[width=0.5\textwidth]{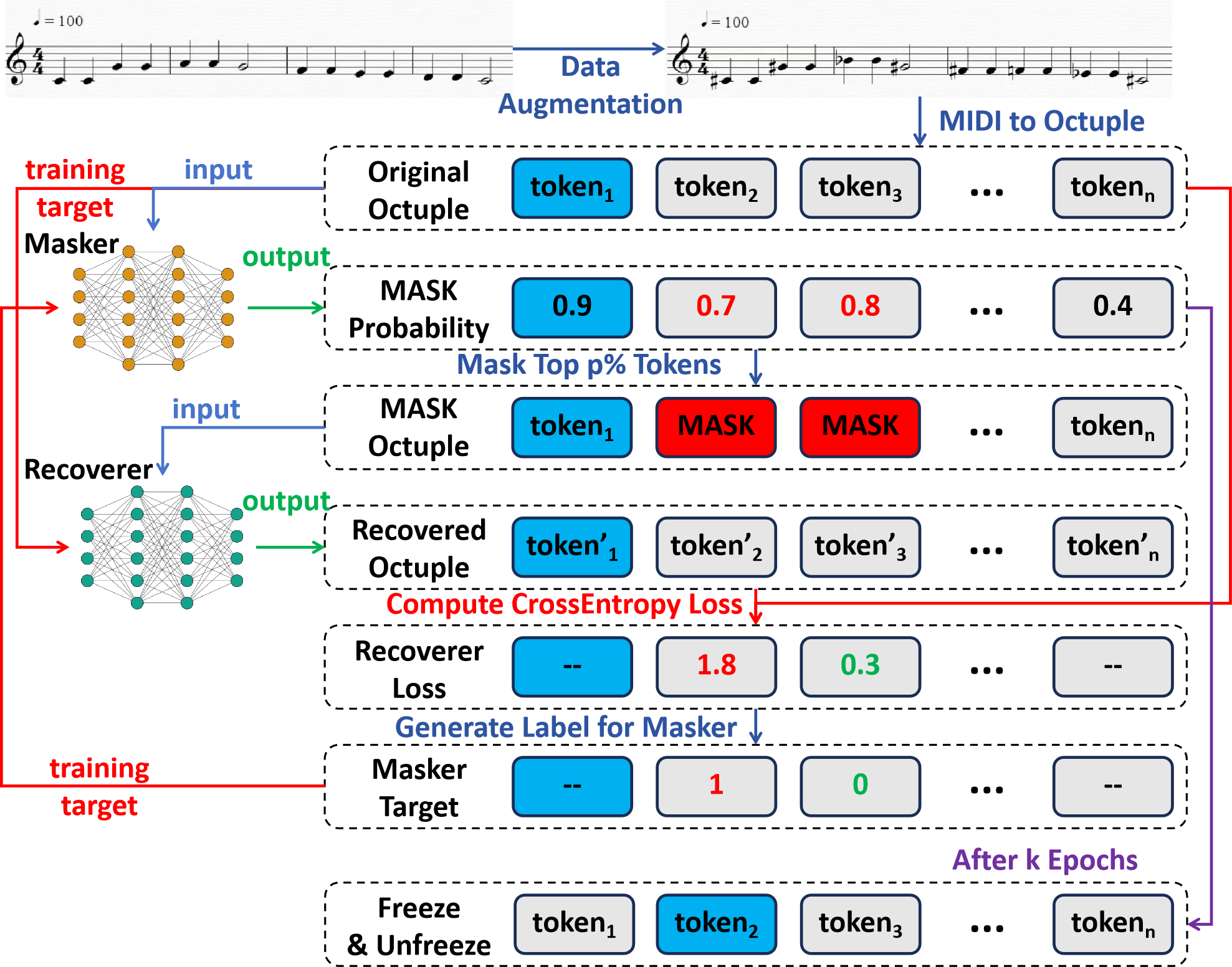}
\caption{Pre-train Process: The blue tokens represent the frozen tokens, which cannot be selected as [MASK] tokens.}
\label{fig:pretrain}
\end{figure}

\subsection{Adversarial Pre-training}
The pre-training process is shown in Fig. \ref{fig:pretrain}. First, we perform random transposition to expand the training data, as music datasets are limited. The transposition operation randomly raises or lowers the entire pitch according to the twelve-tone equal temperament within an octave. The transposition range limitation ensures that the style or emotion of the song is not significantly changed by the shift in pitch register. After that, we convert the MIDI file to an Octuple token sequence as the model input.

Within each epoch, the masker first generates the masking probability of each token, and the tokens with the highest $p\%$ masking probabilities are chosen. We follow a similar method to BERT, using the [MASK] token to replace 80\% of the chosen tokens and random tokens to replace the remaining 20\%. The masked Octuple sequence is then input to the recoverer. We can calculate recovery loss of each masked token according to the following equation:
\begin{equation}
\begin{aligned}
L_i & = \sum_{j=1}^8 w_j \text{CrossEntropy}(\hat{x}_{i,j},x_{i,j}) \ , \\
L_{recoverer} & = \sum_{i \in S}  L_i \ , 
\label{CE}
\end{aligned}
\end{equation}
where $x_{i,j}$ represents the $j^{th}$ attribute of the $i^{th}$ token, $\hat{x}$ represents the recovered token, $w_j$ is the weight of the $j^{th}$ attribute, and $S$ is the set of masked token indices. The recoverer's loss value is the sum of the recovery loss for those masked tokens. We notice that different attributes have varying convergence speeds and performance, so we design a dynamic weight to balance the loss between them. At the beginning of training, $w_1 \sim w_8$ are set equally to 0.125. Then, in the $n^{th}$ epoch, $w_j$ is set as:
\begin{equation}
\begin{aligned}
w_j = \frac{\frac{1}{a_j}}{\sum_{i=1}^8 \frac{1}{a_i}} \ , 
\label{weight}
\end{aligned}
\end{equation}
where $a_i$ is the average recovered accuracy of the $i^{th}$ attribute in the $(n-1)^{th}$ epoch. This way, the recoverer pays more attention to the attributes with lower accuracy.

The recovery loss of each token is also used to generate the learning target of the masker, which aims to lower the recovered accuracy of the recoverer by selecting tokens with high loss values. To achieve this, we set the learning target of the top $q\%$ tokens with the highest loss values as $1$ and the top $q\%$ tokens with the lowest loss values as $0$. The loss function of the masker can be represented as:
\begin{equation}
\begin{aligned}
L_{masker} = \sum_{i \in I_0} \text{BCE} (p_i,0) + \sum_{i \in I_1} \text{BCE} (p_i,1) \ , 
\label{masker}
\end{aligned}
\end{equation}
where $p_i$ is the masking probability generated by the masker for the $i^{th}$ token, and $I_0, I_1$ represent the token index sets with targets set to 0 or 1, respectively.

After repeating this process for $k$ epochs, we believe the tokens with the highest masking probabilities are the most challenging to recover. These tokens correspond to context-free tokens, as they can only be predicted based on the data distribution of the training set, leading to the lowest accuracy. As a result, we freeze the top $a\%$ tokens within each song to avoid them being chosen in the subsequent training, which can be realized by maintaining a dictionary. Simultaneously, we also randomly unfreeze $b\%$ of the frozen tokens to prevent incorrect freezing in the previous step.

\subsection{MASK Fine-tuning}
During fine-tuning, we can still utilize the data augmentation methods employed in pre-training if the downstream tasks are tonality-independent. However, a potential gap may arise since the [MASK] token is present in every epoch during pre-training but is absent in fine-tuning. To address this, we randomly replace $p\%$ of the input tokens with the [MASK] token during fine-tuning. This approach is also similar to the dropout mechanism, which can also help mitigate overfitting.

\section{Experiment}
\subsection{Experiment Setup}

\begin{table}
\caption{Model Configurations}
    \centering
        \begin{tabular}{|c|c|}
        \hline
        \textbf{Configuration} & \textbf{Our Setting} \\
        \hline 
        Input Length     & 1024   \\
        \hline
        Network Layers   & 12   \\
        \hline 
        Hidden Size   & 768    \\
        \hline 
        Inner Linear Size     & 3072   \\
        \hline 
        Attn. Heads   & 12   \\
        \hline 
        Dropout Rate  & 0.1    \\
        \hline 
        Optimizer    & AdamW \\
        \hline 
        \multirow{2}{*}{Learning Rate}  & $10^{-4}$ (pre-train) \\
        & $10^{-5}$ (fine-tune)  \\
        \hline 
        Batch Size   & 8  \\
        \hline 
        Parameters Mentioned in   & \multirow{2}{*}{(15,30,30,10,15)}  \\
        Section \ref{Proposed Method} $(p,q,a,b,k)$ & \\
        \hline 
        Total Number of Parameters  & 115 Million   \\
        \hline 
        \end{tabular}
\label{tab:configuration}
\end{table}

\begin{table*}[!t]
\caption{Dataset Decription}
    \centering
        \begin{tabular}{|c|c|c|c|c|c|}
        \hline
        \textbf{Dataset} & \textbf{Pieces} & \textbf{Task}  & \textbf{Task Level} & \textbf{Class Number} & \textbf{Used in Pre-training} \\
        \hline 
        ASAP \cite{ASAP} & 1068 & -- & -- & --  & $\checkmark$ \\
        \hline
        Pop1K7 \cite{Pop1K7} & 1747 & -- & -- & --  & $\checkmark$ \\
        \hline
        Pianist8 \cite{Pianist8} & 865  & Composer Classification & Sequence Level & 8 & $\checkmark$ \\
        \hline
        EMOPIA \cite{EMOPIA} & 1078 & Emotion Recognition & Sequence Level & 4 & $\checkmark$ \\
        \hline
        POP909 \cite{POP909} & 909 & Melody Extraction & Token Level & 3 & $\checkmark$ \\
        \hline
        GiantMIDI \cite{GiantMidi}  & 10855 & Velocity Prediction & Token Level & 6 & $\times$ \\
        \hline
        \end{tabular}
\label{tab:dataset}
\end{table*}

\begin{table*}[!t]
    \centering
    \caption{Model Performance in Different Tasks: The bold and underlined value indicates the best and second best result within each task. (Note that the training times provided are for reference only, as the server is shared with other users.)}
    \begin{adjustbox}{width=\textwidth}
        \begin{tabular}{|c||c|c|c||c|c||c|c|}
        \hline
        \multirow{2}{*}{\textbf{Model}} & \multicolumn{3}{c||}{\textbf{Pre-train}} & \multicolumn{2}{c||}{\textbf{Sequence-Level Classification}}  & \multicolumn{2}{c|}{\textbf{Token-Level Classification}}\\
        \cline{2-8}
         & \textbf{Accuracy} & \textbf{Epochs} & \textbf{Time} & \textbf{Composer} & \textbf{Emotion}  & \textbf{Velocity} & \textbf{Melody}\\
        \hline 
        \textbf{MidiBERT \cite{MidiBERT}} & 79.60\% & 500 & 6.44d & 79.07\% & 67.59\% & 44.88\% & 92.53\% \\
        \hline
        \textbf{MusicBERT-QM \cite{QM}} & 80.57\% & 500 & 9.47d & 83.72\% & 69.52\% & \underline{46.71\%} & \underline{92.64\%} \\
        \hline
        \textbf{MusicBERT \cite{MusicBERT}} & 76.01\% & 500 & 10.06d & 86.05\% & 71.06\% & 38.79\% & 92.47\% \\
        \hline
        \textbf{PianoBART \cite{PianoBART}} & \textbf{96.67\%} & \textbf{268} & \textbf{3.19d} & 88.37\% & 73.15\% & \textbf{49.37\%} & 92.62\% \\
        \hline \hline
        \textbf{Adversarial-MidiBERT (ours)} & 81.47\% &  \underline{436} & 9.82d & \textbf{97.92\%} & \textbf{79.46\%} & 45.58\% & \textbf{92.68\%} \\
        \hline
        \textbf{Adversarial-MidiBERT (fine-tune w/o mask)} & 81.47\% & \underline{436} & 9.82d & 65.98\% & 70.53\% & 45.30\% & 92.55\% \\
        \hline
        \textbf{Adversarial-MidiBERT (pre-train w/o adversary)} & \underline{83.51\%} & 500 &  \underline{5.93d} & \underline{88.91\%} & \underline{74.07\%} & 45.44\% & \underline{92.64\%} \\
        \hline
        \textbf{Adversarial-MidiBERT (w/o pre-train)} & -- & -- & --  & 79.76\% & 68.75\% & 38.70\% & 87.98\% \\
        \hline
        \end{tabular}
     \end{adjustbox}

\label{tab:exp}
\end{table*}

Our model configuration is shown in Table \ref{tab:configuration}. We conduct our experiment using two NVIDIA V100 GPUs. During training, we observe that our Adversarial-MidiBERT occupies about 27GB GPU memory. 

The dataset used in this paper is shown in Table \ref{tab:dataset}. We use five public MIDI datasets to train our model. We then conduct four different downstream tasks to evaluate our model's performance, including two token-level classification tasks and two sequence-level tasks:
\vspace{-0.5em}
\begin{itemize}
\item[$\bullet$] \textbf{Composer Classification:} Similar to style classification, composer classification is a more challenging and fine-grained task. It requires the model to identify which composer created the songs.
\item[$\bullet$] Emotion Recognition: The music emotions in EMOPIA \cite{EMOPIA} are divided into four types: HVHA, HVLA, LVHA, and LVLA. This task requires the model to classify each song into one of these types.
\item[$\bullet$] \textbf{Melody Extraction:} Each song has different sections, including melody, bridge, and accompaniment. This task requires the model to identify which paragraph each token belongs to.
\item[$\bullet$] \textbf{Velocity Prediction:} Since many MIDI files do not include velocity information, it is important to predict the velocity. We divide velocity into six types and train the model to predict it. To avoid information leakage, we use GiantMIDI \cite{GiantMidi}, which does not participate in pre-training and has its velocity information masked. Since our device could not support training using the full dataset, we select only the first 1000 pieces for the experiment.
\end{itemize}
We split each dataset into 80\% training set, 10\% validation set, and 10\% testing set. We employ the same early stopping strategy as \cite{PianoBART,MidiBERT}, where the training would stop if the model's accuracy does not increase on the validation set for 30 consecutive epochs. We also set the maximum training epochs to 500.

\subsection{Experiment Result}

In this section, we compare our method with other SMU models, including MidiBERT \cite{MidiBERT}, MusicBERT \cite{MusicBERT}, MusicBERT-QM \cite{QM}, and PianoBART \cite{PianoBART}. The main differences between these methods lie in their pre-training approaches. For fairness, these BERT-based models use the same backbone as our method. The experimental results are shown in Table \ref{tab:exp}. It can be seen that our method outperforms the previous BERT-based methods in most tasks, but loses to PianoBART in pre-training and velocity prediction. This may be influenced by the model structure, as the auto-regressive mechanism of the decoder structure in Bidirectional and Auto-Regressive Transformers (BART) \cite{BART} makes it more suitable for token-level tasks. However, it can be noticed that our model has an extremely significant increase in performance on sequence-level tasks. What's more, we also conducted an series of ablation study to illustrate our method's performance without pre-training, pre-training without the adversarial-learning mechanism and without masking during fine-tuning. The results show that the model's performance decreases to varying degrees in these cases, demonstrating the effectiveness of our proposed mechanisms. 


\section{Conclusion}
In this paper, we present Adversarial-MidiBERT for SMU, addressing the bias problem of pre-trained models in MIR through an adversarial pre-training method. Additionally, we introduce a mask fine-tuning approach that significantly enhances the model's accuracy on downstream tasks. Our method achieves remarkable performance on four SMU tasks, particularly on sequence-level tasks. In the future, we aim to explore the application of our method to music generation and NLP tasks. Furthermore, additional experiments can be conducted to further elucidate the principles of the proposed adversarial mechanism.
\clearpage

\bibliographystyle{ACM-Reference-Format}
\bibliography{sample-base}

\end{document}